\documentclass[12pt]{article}
\date{}
\title{Neutrino mass generation in the SO(4) model}
\author{Indranath Bhattacharyya\\
Department of Mathematics\\
Govt. College of Engineering and Ceramic Technology,\\
73 A.C. Banerjee Lane, Kolkata-700 010, INDIA\\E-mail :
$i_{-}bhattacharyya@hotmail.com$\\} \vskip .1in
\begin{document}
\maketitle \baselineskip .3in \noindent
\begin{abstract}
Generation of neutrino mass in $SO(4)$ model is proposed here. The
algebraic structure of $SO(4)$ is same as to that of
$SU(2)_{L}\times SU(2)_{R}$. It is shown that the spontaneous
symmetry breaking results three massive as well as three massless
gauge bosons. The standard model theory according to which there
exist three massive gauge bosons and a massless one is emerged
from this model. In the framework of $SU(2)_{L}\times SU(2)_{R}$ a
small Dirac neutrino mass is derived. It is also shown that such
mass term may vanish with a special choice. The Majorana mass term
is not considered here and thus in this model the neutrino mass
does not follow seesaw structure.\vspace{1cm}\\PACs: 11.15.Ex;
11.30.Qc; 12.60.Cn; 14.60.Pq; 14.70.Pw\vspace{0.2cm}
\end{abstract}
\pagebreak
\section{Introduction:} In the framework of standard model the
neutrino is considered as the massless particle. With this
motivation it is also been assumed that no right handed neutrino
can enter in the theory. The situation became changed when
Pontecorvo proposed the neutrino oscillations leading to the
nonzero neutrino mass \cite{Pontecorvo}. After that the solar
neutrino problem and atmospheric neutrino anomaly led to the
concept of the existence of small non-zero neutrino mass. The
neutrino mass physics took a revolutionary shape when at the {\it
Neutrino'98} (June, 1998) conference in Takayama, Japan, the
Super-Kamiokande collaboration announced the discovery of
oscillations of the cosmic ray neutrinos, which would clearly
indicate the existence of neutrino mass. To explain the mass
generation mechanism of the neutrino the standard model theory is
needed to be extended. There are a number of such extension models
incorporating the neutrino mass. There exist two classes of models
where introduction of lepton number violating interactions leads
to radiative generation of small neutrino mass. The first one is
the Zee model \cite{Zee} where a charged $SU(2)_{L}$ singlet field
$\eta^{+}$ is added to the standard model along with a second
Higgs doublet. In the second model, called Babu model \cite{Babu},
one charged field $\eta^{+}$ along with a doubly charged field
$h^{++}$ are added to the standard model, but no second Higgs
doublet is introduced. There is an another model \cite{Mohapatra1}
where the neutrino mass is generated without including the right
handed neutrino in the theory. In this model an additional
$SU(2)_{L}$ triplet Higgs field ($\triangle^{++}, \triangle^{+},
\triangle^{0}$) is added and it breaks the lepton number by two
units and lead to the Majorana mass \cite{Majorana} of the
neutrinos. In fact, the question is not only how to generate the
neutrino mass in the extension of standard model, but also how to
fix the smallness of neutrino mass compared to the mass of the
charged fermions. Another big question arises that the standard
model without right handed neutrino has got the desirable property
that anomaly cancellations imply charge quatization
\cite{Golowich-Pal} if there is only one family of fermions. This
property is lost in presence of right handed neutrinos, as well as
of more than one fermion generation. It has been pointed out
\cite{Babu-Mohapatra} that if the neutrino is considered as
Majorana particle then anomaly cancellation would imply charge
quantization regardless of the number of generations.\\\indent An
elegant way to generate the neutrino mass is to include the right
handed neutrinos in the model which leads to the left right
symmetric model \cite{Pati-Salam, Mohapatra-Pati,
Senjanovic-Mohapatra}. In the standard model $B-L$ is a global
symmetry and cannot be gauged. But when the right handed neutrino
is included in the model the $B-L$ becomes gaugable and the
$SU(2)_{L}\times SU(2)_{R}\times U(1)_{B-L}$ is the gauge group of
the left right symmetric model \cite{Brahmachari,
Mohapatra-marshak}. The seesaw structure of neutrino mass
\cite{Mohapatra-Senjanovic} emerges in this left-right symmetric
model. In this method the neutrino mass is generated by the
spontaneous breaking of global $B-L$ symmetry. It has also been
proposed that the generation of Majorana neutrino mass may be a
consequence of the creation of massless Goldstone boson, called
majoron \cite{Mohapatra-Chikashige}. The neutrinoless double beta
decay process is supposed to emit the majoron, although such
process is subjected to be verified. several experiments have
searched for neutrinoless double beta decay, but clear evidence is
yet to be found. This majoron can couple to the charged fermions.
The upper bound of the strength of Majoron-electron coupling is
calculated in the astrophysical consideration (according to
standard model of sun) as $g_{eJ}\leq 10^{-10}$
\cite{Chanda}.\\\indent In the present article we have proposed a
new mechanism to generate the neutrino mass, which would be simply
the Dirac mass term. We have shown here that in the framework of
$SO(4)$ model the Dirac mass term of neutrino might be very small
without any introduction of Majorana mass term. Hierarchy of the
masses in the fermionic sector could be explained well in this
$SO(4)$ model. The $SO(4)$ model incorporates both left as well as
right hand fields. In Sec-2 we have designed the model to rescue
the standard model after the spontaneous symmetry breaking. In
Sec-3 we have discussed the mass generation in the fermionic
sector and shown how to generate neutrino mass using this left
right symmetric model.
\section{Mass generation of Gauge bosons:}
It is well known that the $SO(4)$ and $SU(2)\times SU(2)$ have the
same Lie Algebraic structures. Therefore, to consider the
left-right symmetric $SU(2)_{L}\times SU(2)_{R}$ model we can
consider simply the $SO(4)$ model. Such model can be thought of as
a generalization of the standard model of the electro-weak theory.
It is quite reasonable to construct the Lagrangian as,
$$\mathcal{L}=\frac{1}{2}(\partial_{\mu}\sigma)^{2}+\frac{1}{2}
\sum_{i=1}^{3}(\partial_{\mu}\phi_{i})^{2}-\mu^{2}[\sigma^{2}+\sum_{i=1}^{3}\phi^{2}]
-\frac{\lambda}{4}[\sigma^{2}+\sum_{i=1}^{3}\phi^{2}]^{2}\eqno{(1)}$$
where,
$$\sigma\rightarrow \sigma+\phi_{i}\overline{\Lambda}_{i}\hspace{1cm}
\phi_{i}\rightarrow\phi_{i}+\varepsilon_{ijk}\Lambda_{j}\phi_{j}-\sigma
\overline{\Lambda}_{i}\eqno{(2)}$$ If we consider a new function
$\Phi$ s.t.
$$\Phi=\left(%
\begin{array}{cc}
  \sigma+i\phi_{3} & \phi_{2}+i\phi_{1} \\
  -\phi_{2}+i\phi_{1} & \sigma-i\phi_{3} \\
\end{array}%
\right)\eqno{(3)}$$ then the Lagrangian takes the form
$$\mathcal{L}=\frac{1}{4}Tr[(\partial_{\mu}\Phi^{\dag})(\partial^{\mu}\Phi)]
-\frac{\mu^{2}}{2}Tr[\Phi^{\dag}\Phi]-\frac{\lambda}{8}(Tr[\Phi^{\dag}\Phi])^{2}\eqno{(4)}$$
Our aim is to generate the mass of the gauge bosons. In the
framework of standard model theory a mass is produced when a gauge
symmetry is spontaneously broken. To make the Lagrangian, given by
the equation (4), a gauge invariant we replace the covariant
derivative $\partial_{\mu}$ by $D_{\mu}$ as follows:
$$D_{\mu}=\partial_{\mu}+g\sum_{\alpha=1}^{3}\frac{\tau_{\alpha}}{2}L_{\mu}^{\alpha}
-g'\sum_{\alpha=1}^{3}\frac{\tau_{\alpha}}{2}R_{\mu}^{\alpha}\eqno{(5)}$$
Here $L_{\mu}^{\alpha}$ and $R_{\mu}^{\alpha}$ are the gauge
fields associated with $SU(2)_{L}$ and $SU(2)_{R}$ respectively.
The $\Phi$ defined by the equation (3) has the transformation
properties as follows:
$$\Phi'=U_{L}\Phi U_{R}^{\dag}\eqno{(6)}$$
where,
$$U_{L}\equiv e^{i\Lambda_{L}^{\kappa}\frac{\tau_{\kappa}}{2}}\hspace{1cm}U_{R}\equiv
e^{i\Lambda_{R}^{\kappa}\frac{\tau_{\kappa}}{2}}$$

We would like to design the model in such a way that it generates
the masses of all gauge bosons, all known fermions and in addition
the neutrino masses. In this model the gauge symmetry is
spontaneously broken when $\Phi$ takes up the vacuum expectation
value, i.e.,
$$\langle\Phi\rangle=v\left(%
\begin{array}{cc}
  1 & 0 \\
  0 & 1 \\
\end{array}%
\right)\eqno{(7)}$$ A small perturbation about
$\langle\Phi\rangle$ generates the gauge boson mass. The
Lagrangian for the mass term is obtained as
$$\mathcal{L}_{mass}=-\frac{v^{2}}{8c_{W}^{2}}\sum_{\alpha}(c_{W}L_{\mu}^{\alpha}
-s_{W}R_{\mu}^{\alpha})^{2}\eqno{(8)}$$ The above Lagrangian shows
that we lead to a situation in which three massive fields
$c_{W}L_{\mu}^{\alpha} -s_{W}R_{\mu}^{\alpha}$ along with three
massless fields $s_{W}L_{\mu}^{\alpha}+c_{W}R_{\mu}^{\alpha}$ are
present. From the equation (8) the Lagrangian for the mass terms
can also be written as
$$\mathcal{L}_{mass}=\mathcal{L}_{mass}^{1}+\mathcal{L}_{mass}^{2}\eqno{(9)}$$
where,
$$\mathcal{L}_{mass}^{1}=-\frac{v^{2}}{8}[(L_{\mu}^{1})^{2}+(L_{\mu}^{2})^{2}]-
\frac{v^{2}}{8c_{W}^{2}}[c_{W}L_{\mu}^{3}-s_{W}R_{\mu}^{3}]^{2}
+0[s_{W}L_{\mu}^{3}+c_{W}R_{\mu}^{3}]^{2}\eqno{(9a)}$$
$$\mathcal{L}_{mass}^{2}=-\frac{v^{2}}{8}[(R_{\mu}^{1})^{2}+(R_{\mu}^{2})^{2}]\eqno{(9b)}$$
Let us watch carefully the terms $\mathcal{L}_{mass}^{1}$ and
$\mathcal{L}_{mass}^{2}$. The terms $\mathcal{L}_{mass}^{1}$ looks
like the SM Lagrangian for mass terms of gauge bosons. Only
difference is that there exists $R_{\mu}^{3}$ field instead of the
field $B_{\mu}$. We know the generator $T_{R}^{3}$ is associated
with $R_{\mu}^{3}$ whereas $B_{\mu}$ corresponds to $Y$. In the
$SU(2)_{L}\times SU(2)_{R}\times U(1)_{B-L}$ framework
Gellmann-Nishigima type of relation
$$Y=2T_{R}^{3}+(B-L)\eqno{(10)}$$
is well known and in this model the Majorona neutrino mass is
generated by the spontaneous breaking of $B-L$ symmetry. But we
don't incorporate the Majorana mass term as $B-L$ maintains a
perfect symmetry in the particle world so far the occurrence of
double beta decay process is established. Therefore in our model
we simply identify
$$Y=2T_{R}^{3}\eqno{(11)}$$
and then the equation (9a) becomes
$$\mathcal{L}_{mass}=-\frac{v^{2}}{8}[(W_{\mu}^{1})^{2}+(W_{\mu}^{2})^{2}]-
\frac{v^{2}}{8c_{W}^{2}}[c_{W}W_{\mu}^{3}-s_{W}B_{\mu}]^{2}
+0[s_{W}W_{\mu}^{3}+c_{W}B_{\mu}]^{2}\eqno{(12)}$$ with taking
$L_{\mu}^{1}=W_{\mu}^{1}$, $L_{\mu}^{2}=W_{\mu}^{2}$,
$L_{\mu}^{3}=W_{\mu}^{3}$ and $R_{\mu}^{3}=B_{\mu}$. Thus the
standard model is rescued by the equation (12), but in addition to
that we get the mass term of $R_{\mu}^{1}$ and $R_{\mu}^{2}$ in
equation (9b) that is beyond the standard model scenario. Even no
such right hand gauge boson fields have been detected
experimentally, although their existence cannot be ruled out. At
least we can say that all the known particles, predicted in the
standard model theory, are incorporated in the $SO(4)$ model.
\section{Mass generation of fermionic sector:}
Let us now consider the fermionic sector. It is quite clear that
the fermion contents in this model are given by
$$\left(%
\begin{array}{c}
  \nu \\
  l \\
\end{array}%
\right)_{L}\equiv
\left(%
\begin{array}{c}
  u \\
  d \\
\end{array}%
\right)_{L}\equiv(2,1)\hspace{2cm}
\left(%
\begin{array}{c}
  \nu \\
  l \\
\end{array}%
\right)_{R}\equiv
\left(%
\begin{array}{c}
  u \\
  d \\
\end{array}%
\right)_{R}\equiv(1,2)$$ where, $\nu, l, u$ and $d$ stand for
neutrino, lepton, up and down quark respectively of all three
generations. \\\indent In the $SO(4)$ model the Yukawa term of the
Lagrangian is taken as
$$\mathcal{L}^{Y}=\sum_{f}g_{f}[\left(%
\begin{array}{cc}
  \overline{u}^{f} & \overline{d}^{f} \\
\end{array}%
\right)_{L}\left(%
\begin{array}{cc}
  v & 0 \\
  0 & v \\
\end{array}%
\right)(1-x_{f}\tau_{2})\left(%
\begin{array}{c}
  u^{f} \\
  d^{f} \\
\end{array}%
\right)_{R}+h.c.]\eqno{(13)}$$ where, $\left(%
\begin{array}{cc}
  u^{f} & d^{f} \\
\end{array}%
\right)$ represents a general fermionic doublet and $x_{f}$
represents a real number lying between -1 to 1. From the above
Lagrangian we obtain the fermionic mass term as
$$\mathcal{L}_{mass}^{Y}=\sum_{f}\frac{g_{f}v}{\sqrt{2}}[(1-x_{f})\overline{u}^{f}u^{f}+(1+x_{f})
\overline{d}^{f}d^{f}]\eqno{(14)}$$ Clearly the difference of the
masses of up and down quark (of all generations) arise due to the
factor $|x_{f}|=\frac{m_{u_{f}}}{m_{d_{f}}}$. Let us now study the
leptonic sector. The mass term generated in the above manner
becomes
$$\mathcal{L}_{mass}^{l}={g_{l}v}{\sqrt{2}}[(1-x_{l})\overline{\nu}\nu+(1+x_{f})
\overline{l}l]\eqno{(15)}$$ [Note that in the equation (15) $\nu$
denotes the neutrino of generation $l$ lepton. For the convenience
of notation the generation index $l$ is simply dropped from the
suffix of $\nu$.]\\\indent Thus the mass of the neutrino becomes
$$m_{\nu_{l}}=\frac{g_{l}v}{\sqrt{2}}(1-x_{l})\eqno{(16)}$$
that is to be identified as the Dirac mass of the neutrino. We see
that in the $SO(4)$ model Dirac mass of neutrino can be produced
without any difficulty. Also in the leptonic sector there is a
scope to recover the standard model theory by assuming $x_{l}=1$
leading zero neutrino mass. Therefore, it is quite clear that
$x_{f}$ plays a key role to fix the mass of quark and lepton
sector for a given generation. Although we cannot establish any
rule to fix the nature of $x_{f}$ for different quarks and
leptons, but at least we can say that for $0<x_{l}\ll 1$ a small
neutrino mass is obtained.
\section{Discussion}
The model that we have discussed here is not only simple, but it
also rescues the standard model of the electroweak theory with
some special choices. If we look on the well known
$SU(2)_{L}\times SU(2)_{R}\times U(1)_{B-L}$ we see that the
Majorana mass term of the neutrino is generated by the spontaneous
breaking of global $B-L$ symmetry and the unbroken symmetry is
$U(1)_{Q}$. Such symmetry breaking occurs in two steps. But in our
model the symmetry is broken once from $SU(2)_{L}\times SU(2)_{R}$
to $\times SU(2)_{R}$, not to any $U(1)$ symmetry. That is the
unbroken symmetry in our model is $\times SU(2)_{R}$. The
generator $Y$ is one of three generators of a $SU(2)$ group and
ultimately we get $Q$ through the Gellmann-Nishigima relation as
in the usual standard model case. However, we cannot detect the
fields associated to the other two generators. Another simplicity
of $SO(4)$ model is the absence of Majorana type of mass. In this
model the Dirac mass term may be very small and thus the smallness
of the neutrino mass can be explained without introducing any
Majorana mass term. The seesaw structure of the neutrino mass is
not emerged in this model. Therefore, the model can be considered
as an alternative of the seesaw mechanism.

\end{document}